\documentclass{icrc29}
\usepackage{unites2e}
\usepackage{graphicx,amssymb,amsmath,times}
\begin{document}
\title[VERITAS Timing]{Exploiting VERITAS Timing Information}
\author[J. Holder et al.] {J. Holder$^a$ for the VERITAS Collaboration$^b$\\
        (a) School of Physics and Astronomy, University of Leeds, U.K.\\ 
        (b) For full author list, see J. Holder's paper ``Status and
  Performance of the First VERITAS Telescope'' in these proceedings.
        }
\presenter{Presenter: J. Holder (jh@ast.leeds.ac.uk), \  
uki-holder-J-abs2-og27-poster}

\maketitle

\begin{abstract}
  The 499 pixel photomultiplier cameras of the VERITAS gamma ray telescopes
  are instrumented with 500MHz sampling Flash ADCs. This paper describes a
  preliminary investigation of the best methods by which to exploit this
  information so as to optimize the signal-to-noise ratio for the detection of
  Cherenkov light pulses. The FADCs also provide unprecedented resolution for
  the study of the timing characteristics of Cherenkov images of cosmic-ray
  and gamma-ray air showers. This capability is discussed, together with the
  implications for gamma-hadron separation.

\end{abstract}

\vspace{-0.3cm}
\section{Introduction}

Imaging atmospheric Cherenkov telescopes use large mirror areas to reflect the
Cherenkov photons from cosmic-ray and gamma-ray air showers onto a
photo-detector camera, usually comprised of photo-multiplier tubes (PMTs). The
number of photoelectrons generated at each PMT is directly proportional
to the charge under the PMT output pulse, and this is easily
measured using ADCs with fixed-length integration gates. An alternative method
is to use ``Flash'' ADCs to rapidly sample the output pulse
and record the complete pulse shape. This allows to maximize the
signal-to-noise ratio for individual pulses at the analysis stage, lowering
the effective energy threshold of the telescope \cite{Buckley03}. A number of
authors have suggested that the timing and pulse shape information could also
be used to improve sensitivity through improved gamma-hadron separation or
better reconstruction of shower parameters (core location, primary energy,
etc.)  \cite{Roberts98, Hess99}.
\vspace{-0.3cm}
\begin{figure}[h]
  \begin{center}
    \begin{tabular}{cc}
      \includegraphics*[height=5.5cm]{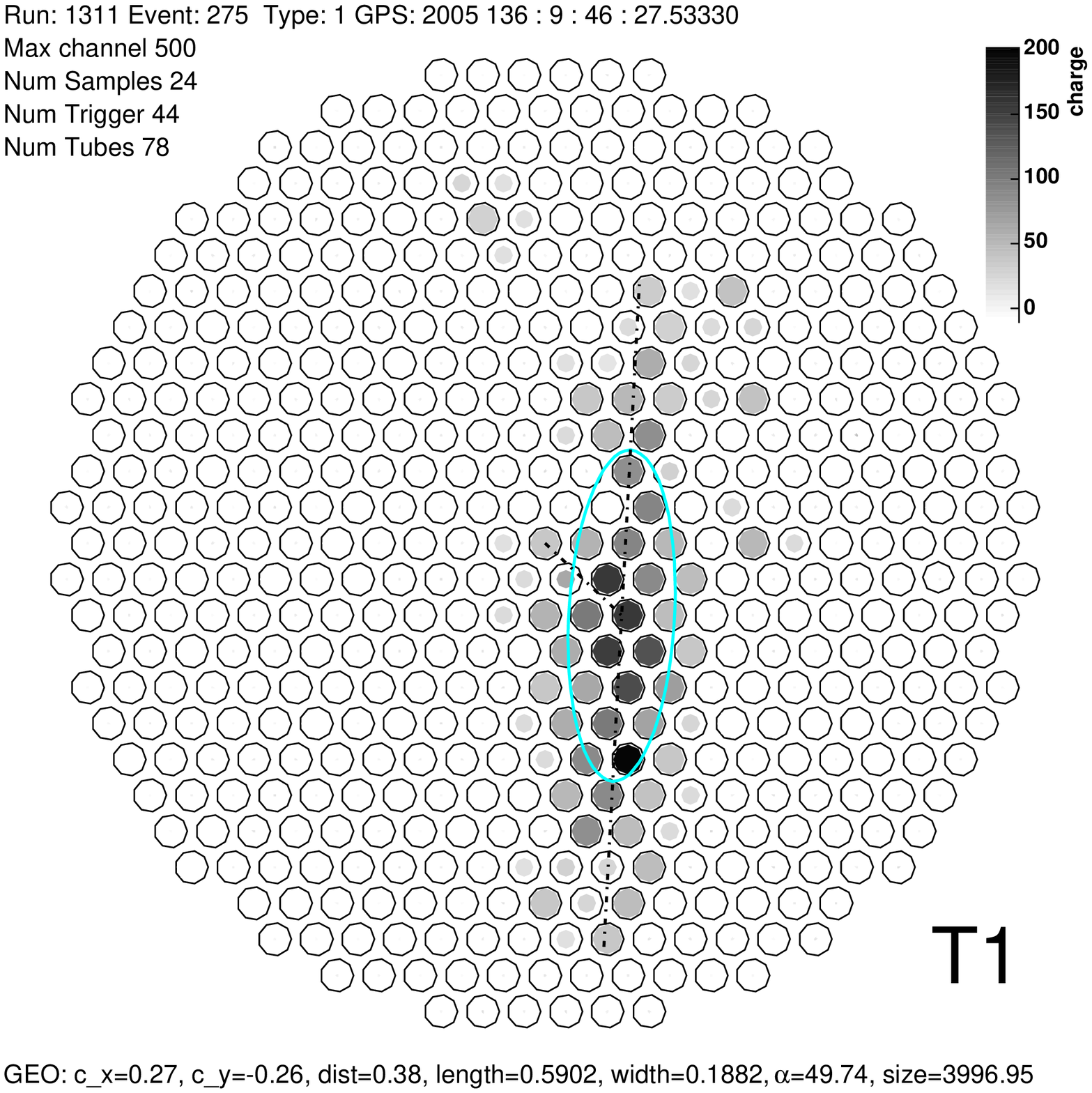}
      \hspace{1cm}
      \includegraphics*[height=5.5cm]{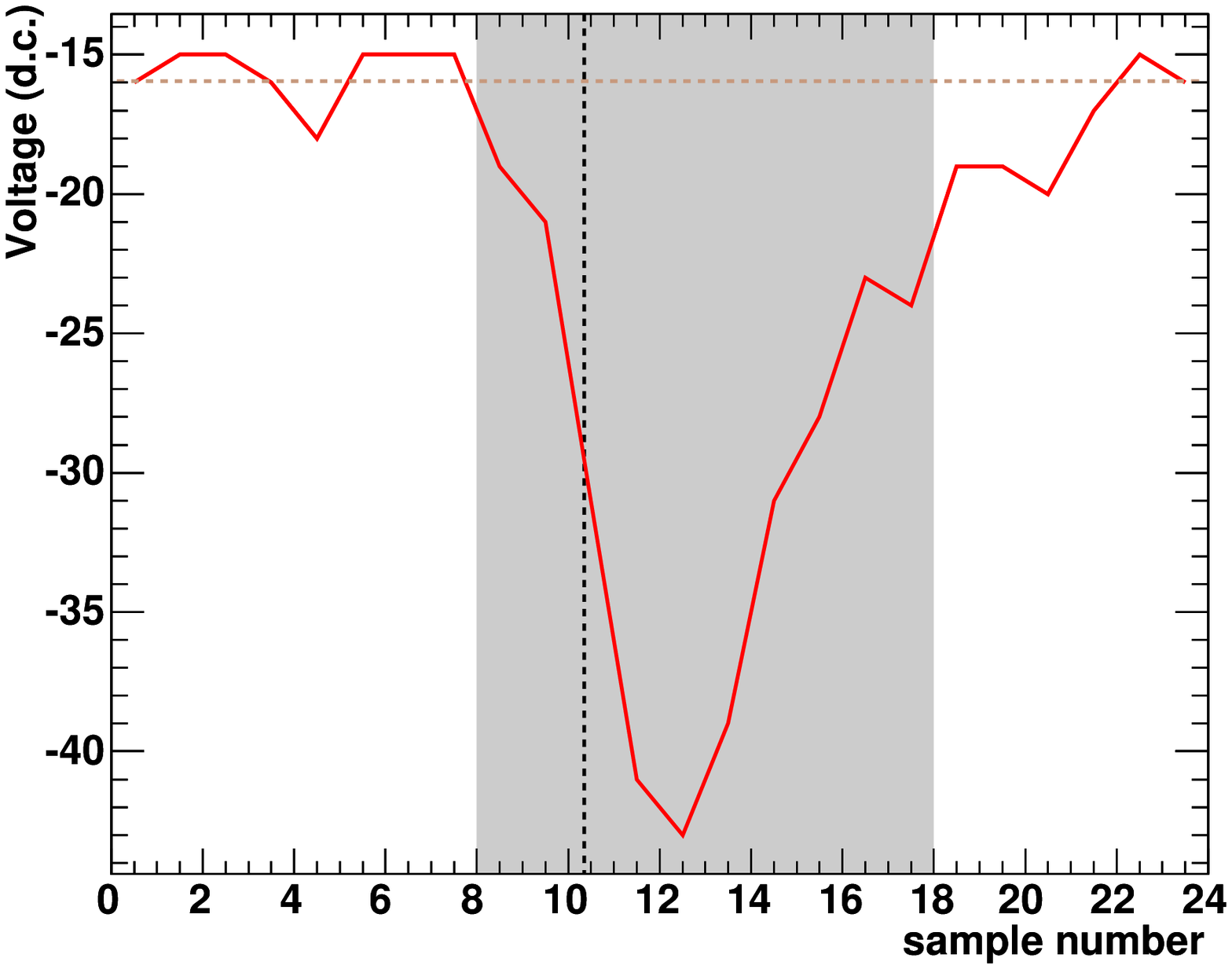}
    \end{tabular}
    \caption{\label {event}
      {\bf Left:} The charge distribution across the camera for a cosmic-ray
      event (the grey scale is in d.c.). {\bf Right:} The FADC trace for a
      PMT in the image. The dashed vertical line indicates $T_0$ for this pulse.
    }
  \end{center}
\end{figure}
\vspace{-0.3cm}

The VERITAS collaboration have been operating the first of four $12\U{m}$
aperture telescopes at the Mt Hopkins basecamp (altitude=$1275\U{m}$) since
January 2005 \cite{Holder05}. The 499 pixel camera is instrumented with
500\U{MHz} sampling FADCs with a memory depth of $32\mu$s and a dynamic range
of 8 bits \cite{Buckley03}. The deep memory simplifies array operation and
will allow us to read out all telescopes when the array is triggered,
including those which do not generate a local trigger themselves. Typically,
the readout window is 24 samples (48\U{ns}) per PMT. Figure~\ref{event} shows
a typical event with a FADC trace. At the FADCs, the PMT signal pulse for an
input delta function has a risetime (10\% to 90\%) of $3.3\U{ns}$ and a width
of $6.5\U{ns}$.
\vspace{-0.3cm}
\section{Timing calibration}
\begin{figure}[h]
  \begin{center}
    \begin{tabular}{cc}
      \includegraphics*[height=5.3cm]{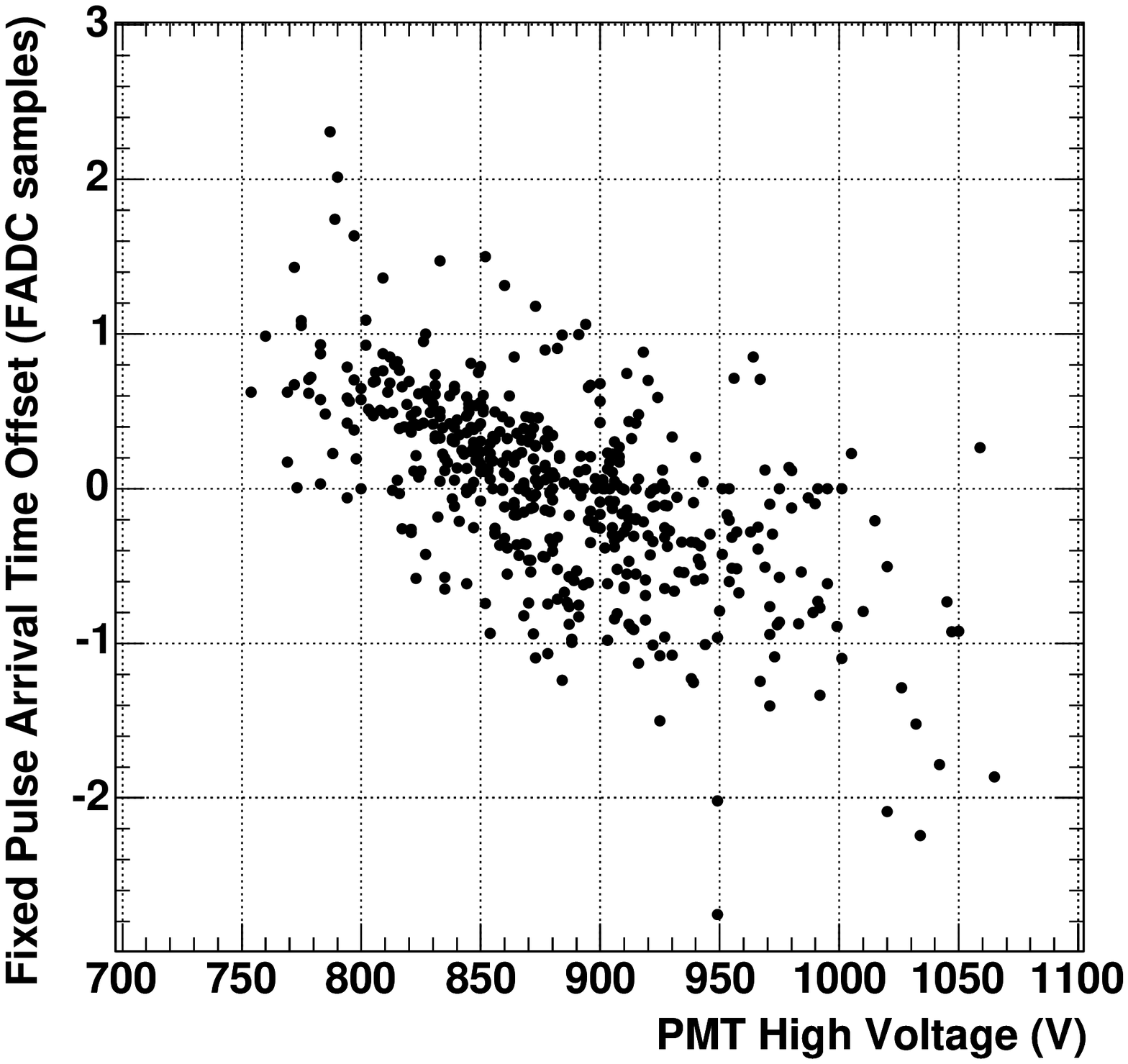}
      \hspace{1cm}
      \includegraphics*[height=5.3cm]{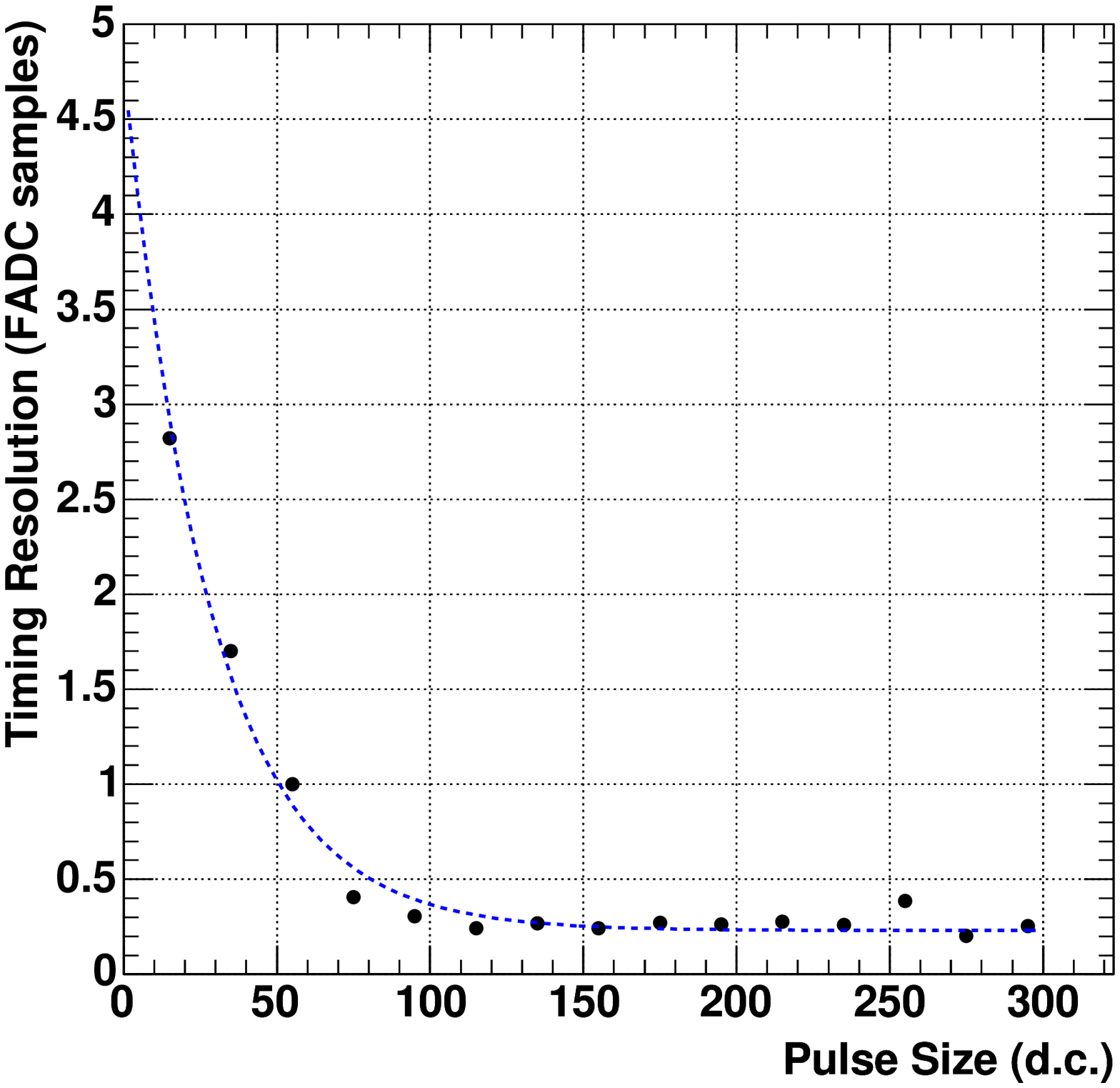}
    \end{tabular}
    \caption{\label {T0dist} {\bf Left:} The mean fixed time offset
    per channel plotted against the high voltage for each PMT. {\bf Right:} The pulse
    timing resolution as a function of the integrated charge in the pulse. The
    dashed line is described in the text.}
  \end{center}
\end{figure}

Before any analysis which uses the pulse timing information can be considered,
the PMT camera timing response must be calibrated. The first stage of this is
to correct for the fixed transit time differences between channels, mainly
caused by the non-uniform high voltages applied to the PMTs. A laser is used
to produce a brief ($\sim4\U{ns}$) light pulse which is sent via fibre-optic
cable to a diffuser situated in front of the PMT camera. At the camera face,
the light flash is uniform in intensity and arrival time. The pulse arrival
time at each PMT, $T_0$, is defined as the half-height point on the pulse
leading edge. The mean value of $T_0$ for a single channel, measured relative
to the mean pulse arrival time over all channels, is the fixed time offset for
that channel. This is subtracted from the measured pulse arrival time in each
channel for each event in order to flat-field the time response over the
camera. Figure~\ref{T0dist} shows these mean values for all channels as a
function of the PMT high voltage. We note that the FADC boards contain a
$6\U{ns}$ programmable delay which is used to compensate for these transit
time offsets at the trigger level to a resolution of $0.85\U{ns}$.
\vspace{-0.3cm}

\vspace{-0.3cm}

The timing resolution, $T_{res}$, for the pulse arrival times is also
important and depends on the pulse time reconstruction method, the pulse
size and the night-sky background level. The VERITAS FADCs are housed in four
VME crates; while the timing between FADCs in the same crates is synchronised,
there is a crate-to-crate jitter of $\sim1$~FADC sample ($2\U{ns}$) which
adversely affects the timing resolution. To correct for this, one
FADC channel in each crate receives a copy of the telescope trigger
signal. This acts as a fixed time reference for each crate which can be
used to correct the crate-to-crate jitter on an
event-by-event-basis. Figure~\ref{T0dist} shows the time resolution for one
channel as a function of laser pulse size. The empirically derived function
shown is $T_{res}=13\exp^{-0.035*(size+30)}+T_{res0}$, where $size$ is the
integrated charge in the pulse and $T_{res0}$ is the timing resolution for
large pulses, typically a few 100~picoseconds.

\vspace{-0.3cm}
\section{Gamma-Hadron Separation}
The longitudinal development of an air shower is reflected in the long axis of
the elliptical image recorded in the camera. The photon arrival time profile
along this axis is largely a result of geometrical path length differences,
and hence the shower core distance. As the shower particles move faster than
the speed of light in air, when the shower has a small core distance
Cherenkov light emitted from lower in the atmosphere is received at the
telescope first. At large core distances, this situation is reversed, as the
Cherenkov light travel time from the shower to the telescope dominates. The
effect of this is to produce a timing gradient along the long axis of the
image, the size and sign of which depend upon the core distance. For
gamma-ray showers from a point source at the centre of the field of view, the
shower core distance is directly related to the angular $distance$ in the
camera of the image from the source position.

\begin{figure}[h]
  \begin{center}
    \begin{tabular}{cc}
      \includegraphics*[height=5.3cm]{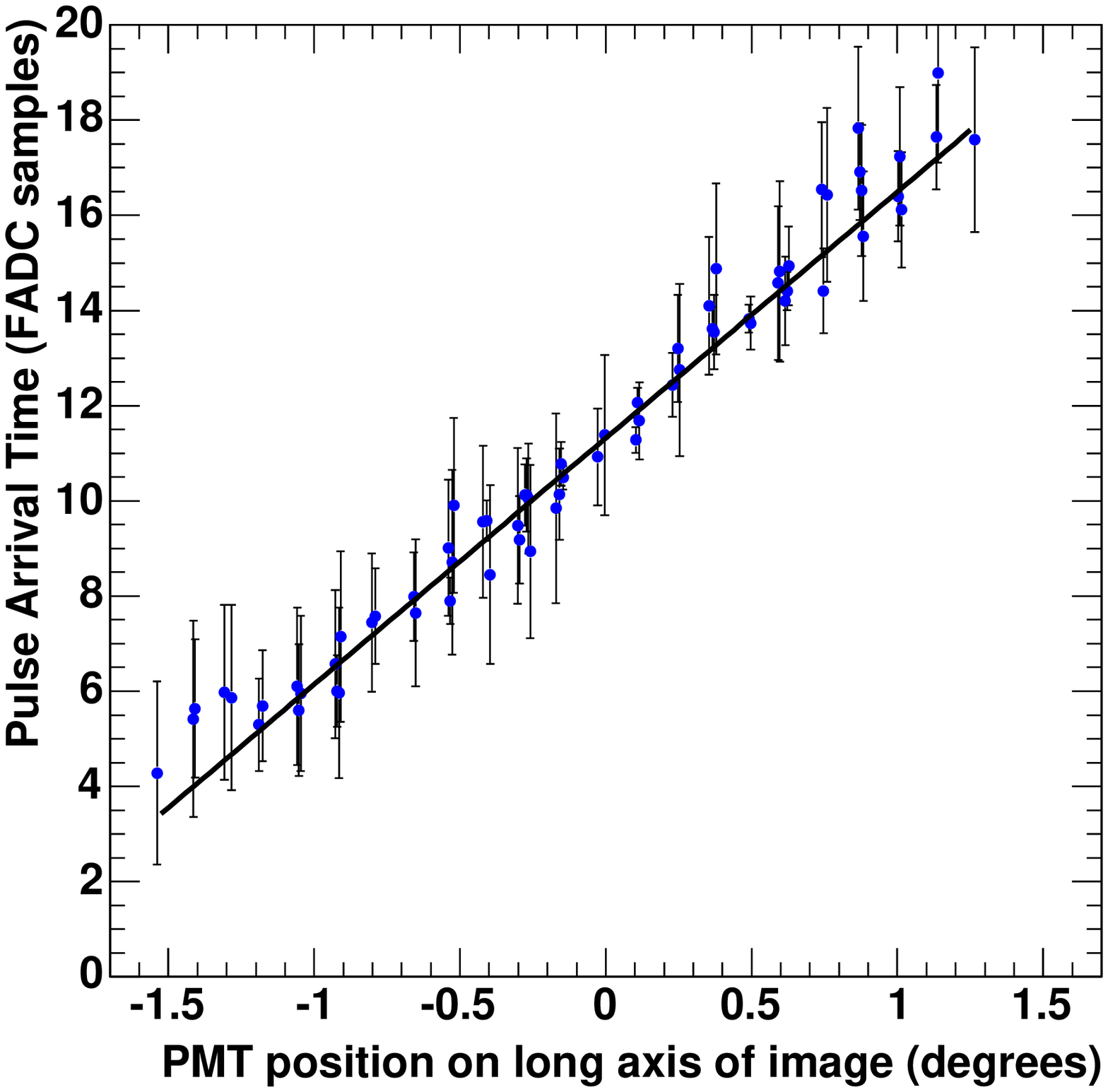}
      \hspace{1cm}
      \includegraphics*[height=5.3cm]{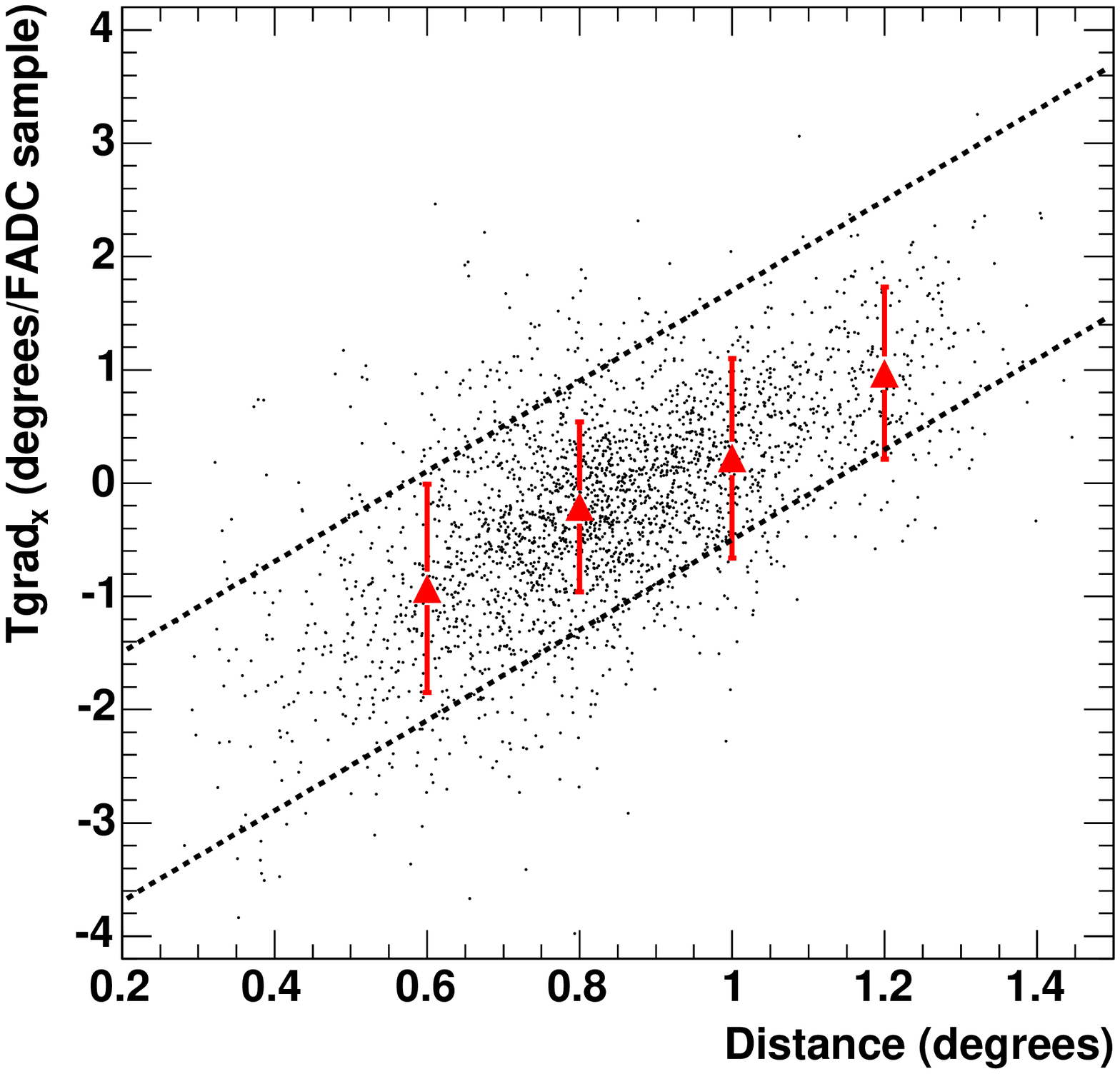}
    \end{tabular}
    \caption{\label {Tgrad_x} {\bf Left:} The Cherenkov pulse arrival time
      distribution (in units of FADC samples = $2\U{ns}$) along the long axis
      of the image in figure~\ref{event}.  {\bf Right:} $Tgrad_x$ as a function
      of $distance$ for real (large triangles) and simulated (small points)
      gamma-ray showers. Dashed lines show the optimum gamma-ray selection region.}
  \end{center}
\end{figure}

Figure~\ref{Tgrad_x} shows the pulse arrival time as a function of PMT
position along the long axis of the image for the cosmic-ray event shown in
figure~\ref{event}. The gradient of a straight line fit to this graph is
$Tgrad_{x}$. Also shown in figure~\ref{Tgrad_x} is the correlation between
$distance$ and $Tgrad_{x}$ for simulated and real gamma-ray showers. The real
data is a combination of high elevation observations of the Crab Nebula and
Markarian 421 with the first VERITAS telescope. Background cosmic-ray images
(and in particular short arcs generated by local muons) which survive the
usual gamma-ray cuts have a flat distribution of $Tgrad_x$, with a mean value
of $Tgrad_x\sim0$ regardless of the image $distance$. This implies that it may
be possible to use the image timing gradient as an additional gamma-hadron
discriminant. We do this by fitting a line to the data points in
figure~\ref{Tgrad_x} and then selecting a gamma-ray region defined by an upper
and lower bound parallel to this line. The dashed lines in the figure show the
gamma-ray selection region which provides the most improvement to the
significance; however this improvement is only 5\%.
\vspace{-0.3cm}

\vspace{-0.4cm}
\section{Improving the Signal-to-Noise}

Having detailed pulse shape information also allows us to optimize the
accuracy of the charge measurement by application of signal processing
algorithms. We describe one such method here; various other techniques
(matched filters \cite{Buckley03}, pulse shape fitting, etc.) are also under
investigation.

\begin{figure}[h]
  \begin{center}
    \begin{tabular}{cc}
      \includegraphics*[height=5.3cm]{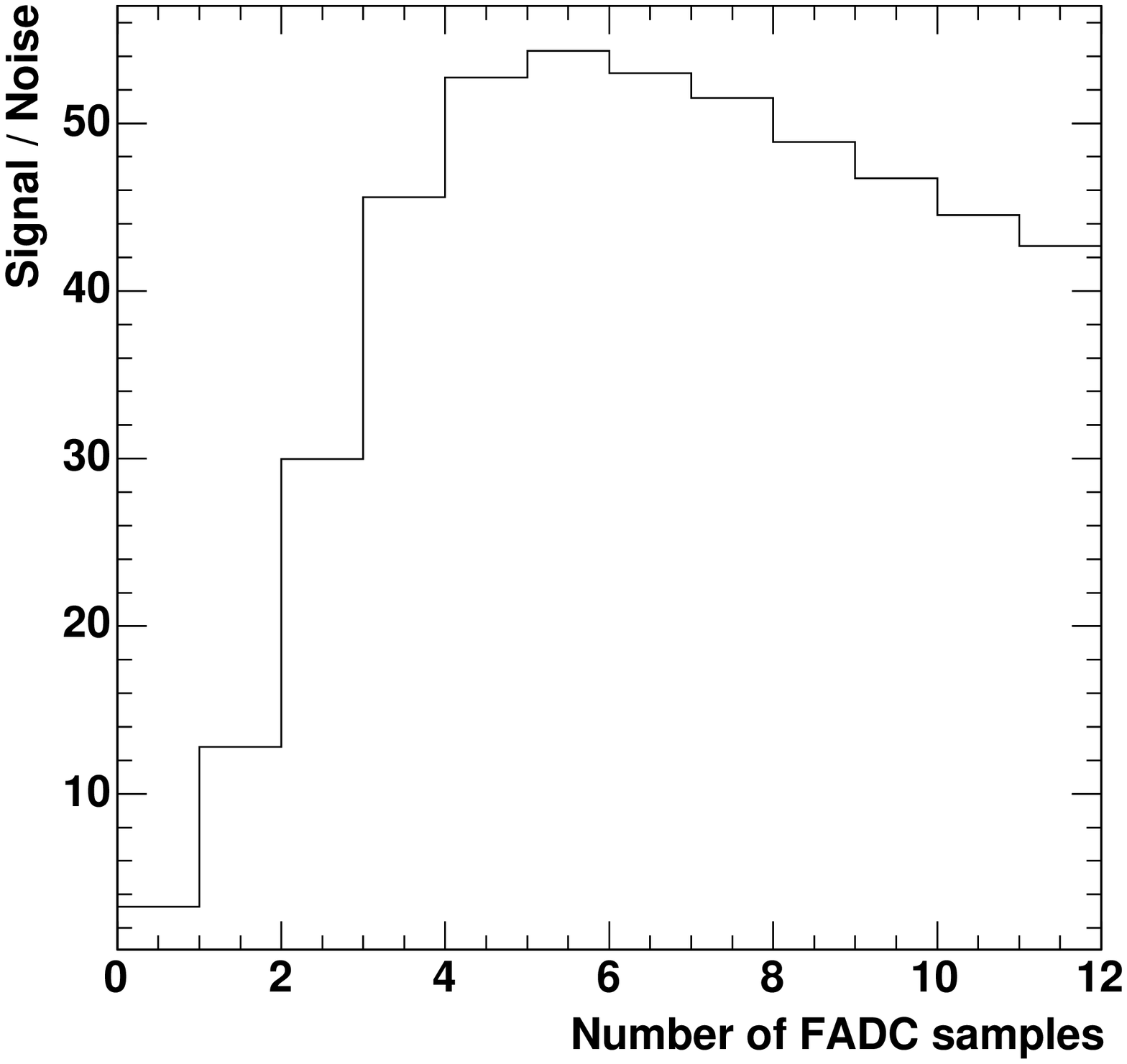}
      \hspace{1cm}
      \includegraphics*[height=5.3cm]{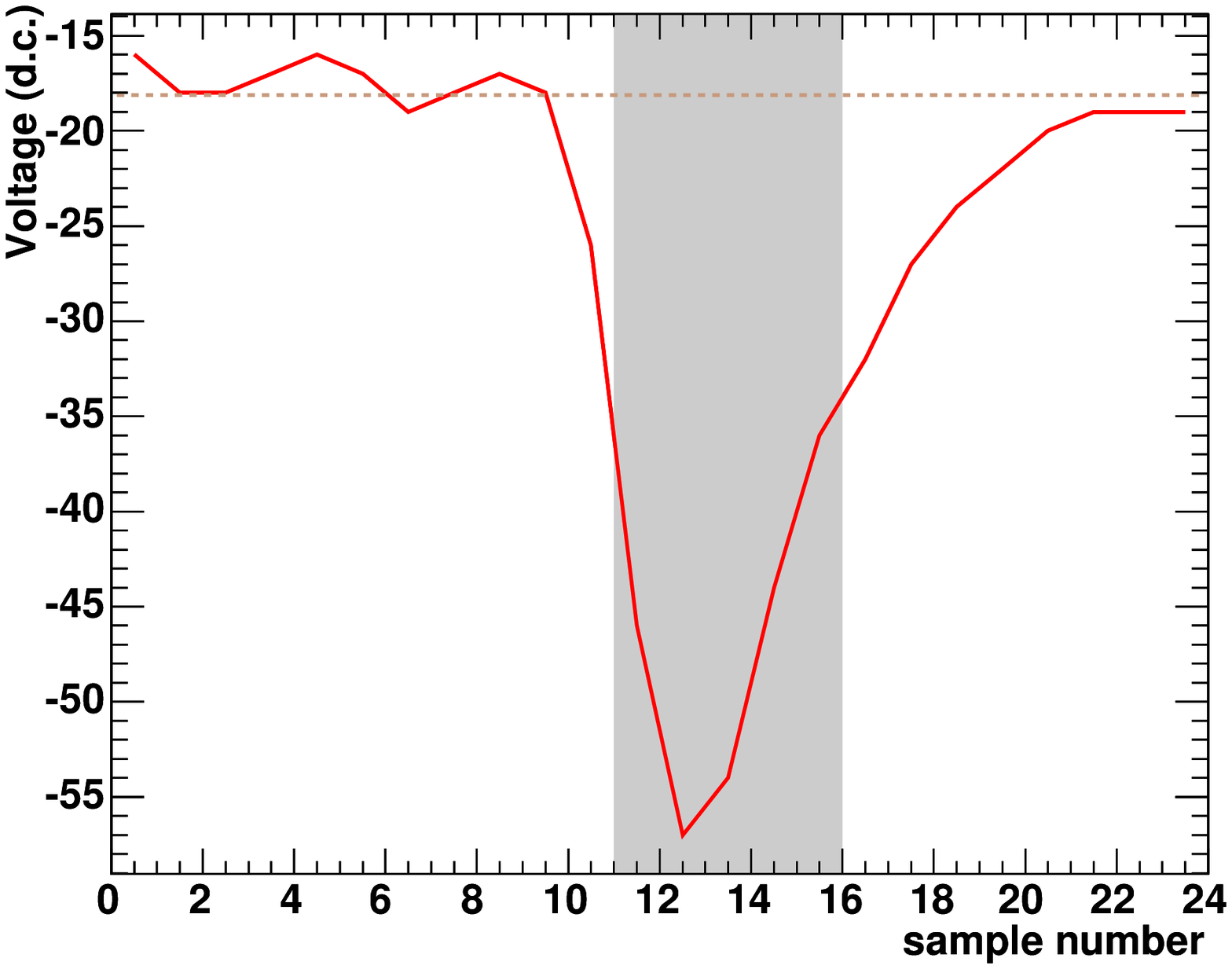}
    \end{tabular}
    \caption{\label {SN} {\bf Left:} The signal-to-noise ratio as a function
      of the integration window size (in FADC samples) {\bf Right:} The FADC
      trace for a PMT with the shaded window indicating the shortened
      integration gate placed according to the image's $Tgrad_x$ value. }
  \end{center}
\end{figure}

The simplest way to improve the signal-to-noise ratio on the charge
measurement is to reduce the size of the FADC integration window; however,
care must be taken that the reduced window is correctly located around the
pulse position. We achieve this by a two-pass method: first, the images are
parameterized using a fixed, wide ($20\U{ns}$) integration gate and $Tgrad_x$
is measured, then the $Tgrad_x$ value is used to define the start position of
a shortened integration gate which will be different for each PMT channel. The
images are then re-parameterized using the shortened gate and the parameters
written out for use in the gamma-ray selection analysis. The optimum size of
the shortened gate is determined to be 5~FADC samples ($10\U{ns}$) by
measuring the ratio of the integrated PMT signal pulse and the night sky
background noise as a function of the integration window size (see figure~\ref{SN}).

The benefit of the two-pass method is not simple to quantify, as other
parameters in the analysis also require optimization (e.g. image cleaning
thresholds, image parameter cut positions), and the best values of these
parameters will be different for the two-pass method and a simple fixed
integration gate method; however the results of applying optimizing image
parameter cuts to a $3.9\U{hour}$ Crab Nebula data set indicate an improvement
to the final significance of 10\%. Perhaps more importantly, this method is
relatively robust against long-term drifts in the trigger time which affect
the average pulse position in the FADC window.
\vspace{-0.3cm}

\vspace{-0.6cm}
\section{Conclusion}
A preliminary examination of the timing information provided by the VERITAS
FADCs verifies that the effective charge integration gate can be reduced to
$10\U{ns}$. While the muon background currently limits the analysis threshold
of the first VERITAS telescope, the reduced charge integration gate should
result in a correspondingly reduced energy threshold for analysis when stereo
data becomes available. This preliminary study does not indicate a significant
improvement in gamma-hadron separation using the timing information; however,
this situation may also change when the muon background is
removed. Refinements to the simulations, the application of advanced digital
signal processing methods and an increased dataset of gamma-ray source
observations should all help future studies in this area.

\end{document}